\documentclass[12pt]{article}

\usepackage{bm}
\usepackage[usenames,dvipsnames]{xcolor}
\usepackage{srcltx}
\usepackage{graphicx}
\usepackage{epstopdf}
\usepackage{color}
\usepackage{amsmath}
\usepackage{amssymb}
\usepackage{color}
\usepackage{float}
\usepackage{subfigure}
\usepackage{gensymb}
\usepackage{appendix}

\usepackage{epstopdf}
\usepackage{epsfig}
\usepackage{psfrag}
\usepackage[mode=errorstop]{pstool}
    \newcommand{\figref}{Fig.~\ref}
  	
 	\newcommand{\secref}{Sec.~\ref}

\begin{document}

\title{Magnetless Reflective Gyrotropic \\ Spatial Isolator Metasurface}


\author{Guillaume Lavigne and Christophe Caloz}

\maketitle


\begin{abstract}
We present the concept of a magnetless Reflective Gyrotropic Spatial Isolator (RGSI) metasurface. This is a birefringent metasurface that reflects vertically polarized incident waves into a horizontally polarized waves, and absorbs horizontally polarized incident waves, hence providing isolation between the two orthogonal polarization. We first synthesize the metasurface using surface susceptibility-based Generalized Sheet Transition Conditions~(GSTCs). We then propose a mirror-backed metaparticle implementation of this metasurface, where transistor-loaded resonators provide the desired magnetless nonreciprocal response. Finally, we demonstrate the metasurface by full-wave simulation results. The proposed RGSI metasurface may be used in various electromagnetic applications, and may also serve as a step towards more sophisticated magnetless nonreciprocal metasurface systems.
\end{abstract}

\section{Introduction}\label{sec:intro}

Nonreciprocity is a fundamental concept in science and technology~\cite{caloz2018electromagnetic,asadchy2020tutorial}. It allows special operations, such as isolation, circulation, nonreciprocal phase shifting and nonreciprocal gyrotropy, that are crucial in a great variety of applications. In electromagnetics, nonreciprocity is conventionally obtained through the use magnetized materials, such as ferrites~\cite{lax1962microwave} or terbium garnet crystals~\cite{villaverde1978terbium}. However, magnetized materials have severe drawbacks, such as incompatibility with integrated circuit technologies and bulkiness due to the required biasing magnets. Recently, the concept of magnetless nonreciprocity has arisen as a potential solution to these issues~\cite{caloz2018guest}, with the transistor-loaded structures~\cite{popa2007architecture,yuan2009zero,popa2012nonreciprocal,kodera2011artificial,wang2012gyrotropic,sounas2012electromagnetic,kodera2013magnetless,ra2016magnet,taravati2017nonreciprocal,kodera2018unidirectional,ra2020nonreciprocal} and spacetime-modulated systems~\cite{hadad2016breaking,shi2017optical,sounas2013giant,sounas2018angular,wang2018time,shaltout2015time,taravati2020full} being the main practical\footnote{Magnetless nonreciprocity can also be obtained by nonlinearity combined by structural asymmetry~\cite{caloz2018guest,sounas2018nonreciprocity,fernandes2018asymmetric}. However, the related systems are generally unpractical for engineering devices, due major issues such as single excitation at a time, poor transmission and isolation performance, and intensity dependence.} approaches.

Advances in magnetless nonreciprocity have recently been extended to  metasurfaces, where magnetized material technologies would be inapplicable. Metasurfaces have experience spectacular developments over the past decade~\cite{glybovski2016metasurfaces,achouri2018design}. They have been shown to provide unprecedented control over the fundamental properties of electromagnetic waves, such of polarization, reflection, refraction, spin and orbital angular momentum. However, most of the studies on metasurfaces reported so far have focused on reciprocal structures. Introducing nonreciprocity in metasurfaces has the potential to extend conventional nonreciprocal operations such as isolation and circulation, usually applied to guided waves, to spatial wave manipulations, and to lead to novel metasurface-based wave transformations. As in other platforms, the transistor-loaded route for nonreciprocity, compared spacetime-varying systems, has the advantage  in metasurfaces to produce no spurious harmonic and intermodulation frequencies while using the simplest form of biasing, namely a simple DC battery. Transistor-loaded nonreciprocal metasurfaces have been demonstrated realizing nonreciprocal polarization rotators in reflection~\cite{kodera2011artificial} and in transmission~\cite{wang2012gyrotropic}, transmissive isolation using an antenna-circuit-antenna approach~\cite{taravati2017nonreciprocal}, bianisotropic nonreciprocity~\cite{ra2016magnet} and meta-grating reflective circulators~\cite{ra2020nonreciprocal}.

The most fundamental and primary application of nonreciprocity in metasurfaces is probably spatial isolation. Here, we introduce the concept of a reflective isolator metasurface, with a pair of orthogonally-polarized ports coupled by reflective gyrotropy, and demonstrate a corresponding magnetless Reflective Gyrotropic Spatial Isolator (RGSI).


\section{Operation Principle}\label{sec:operation_principle}

\begin{figure}[h!]
  \centering
  \includegraphics[width=0.7\columnwidth]{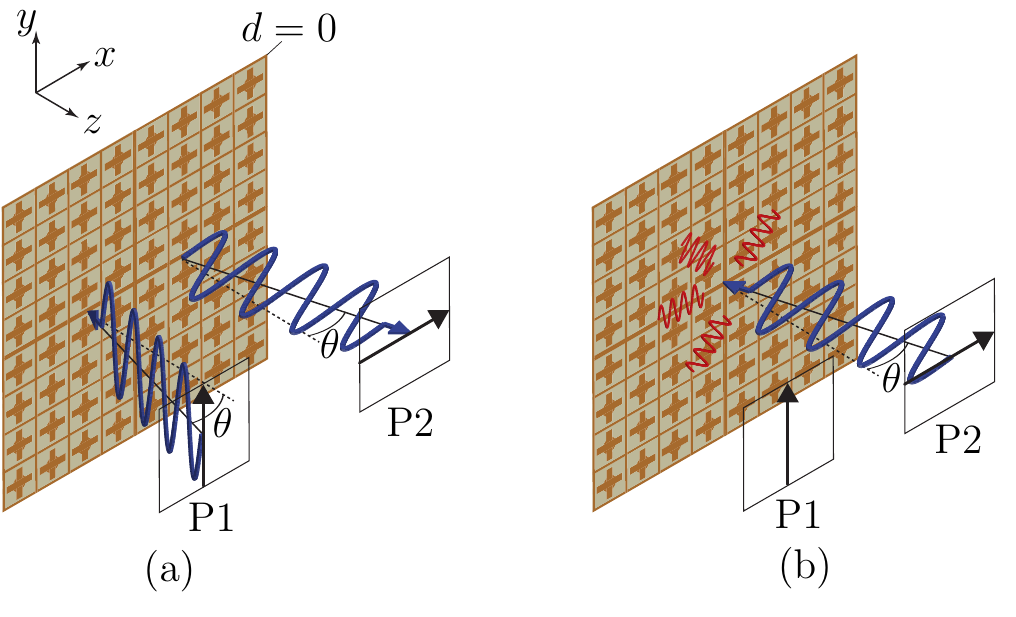}
  \caption{Operation principle of the proposed Reflective Gyrotropic Spatial Isolator (RGSI) metasurface. (a)~A $y$-polarized incident wave, from port P1, is reflected with $x$-polarization, to port P2. (b)~An $x$-polarized incident wave, from port P2, is absorbed by the metasurface.}\label{fig:generic_structure}
\end{figure}

Figure~\ref{fig:generic_structure} depicts the operation principle of the proposed RGSI metasurface. The metasurface includes reciprocity-breaking elements, and is designed in such a manner that, using birefringence, it specularly\footnote{``Specular,'' from the Greek word ``speculum'' that means ``mirror,'' refers to reflection that occurs under the same angle as the incidence angle, according to Snell law of reflection. We restrict here our attention to specular reflection, as implicitly assumed from the equal incidence and reflection angles ($\theta$) in \figref{fig:generic_structure}(a). However, the concept of reflective gyrotropic spatial isolator could naturally be extended to non-specular reflection, with reflection angle differing from the incidence angle, by using metasurface gradient and bianisotropy~\cite{lavigne2018susceptibility}.} reflects vertically-polarized incident waves into horizontally-polarized waves, as shown in \figref{fig:generic_structure}(a), and absorbs horizontally-polarized incident waves, as shown in \figref{fig:generic_structure}(b).

The resulting RGSI device is de facto a two-port reflective spatial isolator, with ports that we denote here P1 and P2, as indicated in the figure. Its electromagnetic response may therefore be described by the following scattering matrix:
\begin{equation}\label{eq:desired_reflection_matrix2}
  \mathbf{S}_\text{spec}^\text{RGSI} =  \begin{bmatrix}
        S_{11}^\text{RGSI} & S_{12}^\text{RGSI} \\
        S_{21}^\text{RGSI} & S_{22}^\text{RGSI}
      \end{bmatrix} =  \begin{bmatrix}
        0 & 0 \\
        A e^{i \phi} & 0
      \end{bmatrix},
\end{equation}
where $A$ and $\phi$ are the amplitude and the phase, respectively, imparted by the metasurface to the rotated reflected-transmitted field. For the polarizations assumed in \figref{fig:generic_structure}, the metasurface may be alternatively described by the reflection matrix

\begin{equation}\label{eq:desired_reflection_matrix}
  \mathbf{R}_\text{spec}^\text{RGSI} = \begin{bmatrix}
        R_{xx}^\text{RGSI} & R_{xy}^\text{RGSI} \\
        R_{yx}^\text{RGSI} & R_{yy}^\text{RGSI}
      \end{bmatrix} =  \begin{bmatrix}
        0 & A e^{i \phi} \\
        0 & 0
      \end{bmatrix},
\end{equation}
so that $\mathbf{R}_\text{spec}^\text{RGSI} = (\mathbf{S}_\text{spec}^\text{RGSI})^\text{T}$.

\section{Metasurface Design}\label{sec:theoretical_derivation}

\subsection{GSTC Equations}

Metasurfaces may be modeled as zero-thickness discontinuities of space via Generalized Sheet Transition Conditions (GSTCs) and bianisotropic surface susceptibility tensors~\cite{achouri2014general,achouri2018design,achouri2020electromagnetic}. The GSTCs, assuming the harmonic time convention $e^{+i\omega{t}}$, read
  \begin{subequations}\label{eq:GSTCs_normal}
  \begin{equation}
    \hat{z} \times \Delta \mathbf{H} = i \omega \mathbf{P} - \hat{z} \times \nabla M_z,
\end{equation}
\begin{equation}
    \hat{z} \times \Delta \mathbf{E} = -i \omega \mathbf{M} - \frac{1}{\epsilon} \hat{z} \times \nabla P_z,
    \end{equation}
  \end{subequations}
where $\Delta\mathbf{H}$ and $\Delta\mathbf{E}$ are the differences of the magnetic or electric fields at both sides of the metasurface, and where $\mathbf{P}$ and $\mathbf{M}$ are the induced electric and magnetic surface polarization densities on the metasurface. The latter may be expressed in terms of surface susceptibility tensors as
  \begin{subequations}\label{polarizabilities_to_susceptibilities}
  \begin{equation}
   \mathbf{P} = \epsilon \overline{\overline{\chi}}_\text{ee} \mathbf{E}_\text{av} + \frac{1}{c} \overline{\overline{\chi}}_\text{em} \mathbf{H}_\text{av},
\end{equation}
\begin{equation}
   \mathbf{M} =  \frac{1}{c}  \overline{\overline{\chi}}_\text{me} \mathbf{E}_\text{av} + \mu \overline{\overline{\chi}}_\text{mm} \mathbf{H}_\text{av},
    \end{equation}
  \end{subequations}
where $\mathbf{E}_\text{av}$ and $\mathbf{H}_\text{av}$ are the averages of the electric or magnetic fields at both sides of the metasurface, and $\overline{\overline{ \chi}}_\text{ee}$, $\overline{\overline{ \chi}}_\text{mm}$, $\overline{\overline{\chi}}_\text{em}$, $\overline{\overline{ \chi}}_\text{me}$ are the $3 \times 3$ bianisotropic susceptibility tensors characterizing the metasurface. In this paper, we shall assume a purely tangential metasurface, i.e., a metasurface with $M_z = P_z =0$, for which the bianisotropic GSTCs simplify to
\begin{subequations}\label{eq:GSTC_tangential}
\begin{equation}
\hat{z} \times \Delta\mathbf{H} = i \omega \epsilon \overline{\overline{ \chi}}_\text{ee} \mathbf{E}_\text{av} +  i k \overline{\overline{ \chi}}_\text{em}   \mathbf{H}_\text{av} ,
\end{equation}
\begin{equation}
\Delta \mathbf{E} \times \hat{z}   = i k \overline{\overline{ \chi}}_\text{me}  \mathbf{E}_\text{av} + i \omega \mu \overline{\overline{ \chi}}_\text{mm} \mathbf{H}_\text{av},
\end{equation}
\end{subequations}
where $\overline{\overline{ \chi}}_\text{ee}$, $\overline{\overline{ \chi}}_\text{mm}$, $\overline{\overline{ \chi}}_\text{em}$, $\overline{\overline{ \chi}}_\text{me}$ are now $2 \times 2$ tensors~\cite{Dehmollaian_TAP2_2019}. In these relations, the differences and averages of the fields are explicitly given by
\begin{subequations}\label{eq:diff_av}
  \begin{equation}
    \Delta \mathbf{\Phi} = \mathbf{\Phi}_\text{t} - (\mathbf{\Phi}_\text{i} + \mathbf{\Phi}_\text{r}),
  \end{equation}
  \begin{equation}
  \mathbf{\Phi}_\text{av} = \mathbf{(\Phi}_\text{t} + \mathbf{\Phi}_\text{i} + \mathbf{\Phi}_\text{r})/2,
  \end{equation}
\end{subequations}
where $\mathbf{\Phi}=\mathbf{E},\mathbf{H}$, where the subscript t, i and r denote the transmitted, incident and reflected fields, respectively.

\subsection{Susceptibility Synthesis}\label{sec:suscepbility_synthesis}

The metasurface can be designed using the susceptibility synthesis procedure described in~\cite{achouri2020electromagnetic}: 1)~specify the desired field transformations, 2)~compute the corresponding field differences and averages, 3)~insert the expressions for these differences and averages into the susceptibility-GSTC equations, and 4)~solve the resulting equations for the surface susceptibility tensors.

The GSTS assumed here, given by~\eqref{eq:GSTC_tangential}, form a linear system of $4$ scalar equations in the $16$ susceptibility components containing the $4$ susceptibility tensors $\overline{\overline{ \chi}}_\text{ee}$, $\overline{\overline{ \chi}}_\text{mm}$, $\overline{\overline{ \chi}}_\text{em}$, and $\overline{\overline{ \chi}}_\text{me}$ of dimensions $2\times{2}$. The isolator operation in \figref{fig:generic_structure} involves 2 non-trivial\footnote{By ``non-trivial'' transformations, we mean here transformations that would not be performed by the simplest metasurfaces, i.e., passive, reciprocal and nongyrotropic metasurfaces.} transformations, specular gyrotropic
reflection-transmission from P1 to P2, and absorption by the metasurface from P2, which implies $2\times{4}=8$ scalar equations in the $16$ susceptibility parameters. This represents an undetermined system, requiring extra specifications for full-rank solvability. Such specifications largely depend from the specific nature of the required transformations.

The transformations in \figref{fig:generic_structure} obviously involve \emph{gyrotropy} and \emph{nonreciprocity}. Nonreciprocity implies $\overline{\overline{\chi}}_\text{ee}\neq\overline{\overline{\chi}}_\text{ee}^T$ or $\overline{\overline{\chi}}_\text{mm}\neq\overline{\overline{\chi}}_\text{mm}^T$ or $\overline{\overline{\chi}}_\text{em}\neq -\overline{\overline{\chi}}_\text{me}^T$~\cite{achouri2020electromagnetic}, where the superscript ``$T$'' denotes the transpose operation, while gyrotropy implies either off-diagonal components of $\overline{\overline{\chi}}_\text{ee}$ and $\overline{\overline{\chi}}_\text{mm}$ or diagonal components of $\overline{\overline{\chi}}_\text{em}$ and $\overline{\overline{\chi}}_\text{me}$~\cite{achouri2020electromagnetic}. This leaves us with several possibilities to eliminate $8$ of the $16$ susceptibility parameters for fully-specified resolution. We choose here, and subsequently implement, a \emph{homoanisotropic design}, characterized by the parameters $\overline{\overline{\chi}}_\text{ee}$ and $\overline{\overline{\chi}}_\text{mm}$ with $\overline{\overline{\chi}}_\text{em}=\overline{\overline{\chi}}_\text{me}=0$, and discuss in Appendix~\ref{sec:bian_des} an alternative \emph{bianisotropic design}\footnote{We follow here the convenient Greek prefix terminology used in~\cite{achouri2020electromagnetic}, where \emph{homo-} involves only the parameters ee and mm, \emph{hetero-} involves only the parameters em and me, and \emph{bi-}, introduced by Kong~\cite{Kong_1972}, involves both homo and bi parameters.}. We are then left with the 8 parameters, with the gyrotropy condition $\chi_\text{ee}^{yx},\chi_\text{ee}^{xy}\neq{0}$ or/and $\chi_\text{mm}^{yx},\chi_\text{mm}^{xy}\neq{0}$ and the
nonreciprocal condition $\chi_\text{ee}^{yx}\neq\chi_\text{ee}^{xy}$ or/and $\chi_\text{mm}^{yx}\neq\chi_\text{mm}^{xy}$, which leads to a full-rank GSTC-susceptibility system.

Considering s-polarization incidence (the p-polarization problem can be treated analogously) and assuming that the metasurface positioned in the plane $z=0$, the $2$ operations in~\figref{fig:generic_structure} correspond to the following tangential field specifications:
\begin{subequations}\label{eq:first_transformation}
\begin{equation}
\mathbf{E}_{i}=   e^{-i k_0 \sin \theta x} \hat{y}, \quad \mathbf{H}_{i}=  e^{-i k_0 \sin\theta x} \cos \theta /\eta \hat{x},
\end{equation}
\begin{equation}
\mathbf{E}_{r}= A e^{-i k_0 \sin \theta x} \cos \theta e^{i \phi} \hat{x}, \quad \mathbf{H}_{r}= A e^{-i k_0 \sin \theta x} e^{i \phi}  /\eta \hat{y},
\end{equation}
\begin{equation}
\mathbf{E}_{t}=0, \quad \mathbf{H}_{t}= 0,
\end{equation}
\end{subequations}
where $\theta$ is the angle of incidence and reflection, for the specular gyrotropic
reflection-transmission from P1 to P2, and
\begin{subequations}\label{eq:second_transformation}
\begin{equation}
\mathbf{E}_{i}= \cos (-\theta) e^{-i k_0 \sin( -\theta) x}\hat{x}, \quad \mathbf{H}_{i}=  -e^{-i k_0 \sin(-\theta) x}/\eta \hat{y},
\end{equation}
\begin{equation}
\mathbf{E}_{r}= 0, \quad \mathbf{H}_{r}=0,
\end{equation}
\begin{equation}
\mathbf{E}_{t}=0, \quad \mathbf{H}_{t}= 0.
\end{equation}
\end{subequations}
and for the absorption by the metasurface from P2.

Substituting the field specifications~\eqref{eq:first_transformation} and ~\eqref{eq:second_transformation} into~\eqref{eq:diff_av}, inserting the resulting expressions into~\eqref{eq:GSTC_tangential}, and solving for the susceptibility tensors yields the sought-after susceptibility synthesis result
\begin{subequations}\label{eq:susceptibilities_full_structure}
  \begin{equation}
    \overline{\overline{\chi}}_\text{ee} = \begin{bmatrix}
                                             \chi_\text{ee}^{xx} & \chi_\text{ee}^{xy} \\
                                             \chi_\text{ee}^{yx} & \chi_\text{ee}^{yy}
                                           \end{bmatrix} = \begin{bmatrix}
                                                             \frac{-2i \sec \theta}{k} & \frac{4i A e^{i \phi}}{k} \\
                                                             0 &  \frac{-2i \cos \theta}{k}
                                                           \end{bmatrix}
  \end{equation}
    \begin{equation}\label{eq:susceptibilities_full_structureb}
    \overline{\overline{\chi}}_\text{mm} = \begin{bmatrix}
                                             \chi_\text{mm}^{xx} & \chi_\text{mm}^{xy} \\
                                             \chi_\text{mm}^{yx} & \chi_\text{mm}^{yy}
                                           \end{bmatrix} = \begin{bmatrix}
                                                             \frac{-2i \sec \theta }{k} & 0 \\
                                                             \frac{4  i A e^{i \phi}}{k} &  \frac{-2i \cos \theta }{k}
                                                           \end{bmatrix},
  \end{equation}
\end{subequations}
where all the components are independent from the spatial variables $x$ and $y$, as might have been expected from the fact that the specified reflection is specular and hence momentum conservative.

\section{Metastructure Implementation}\label{sec:implementation}

\subsection{Metaparticle Configuration}

The next step of the metasurface design is naturally to implement the synthesized susceptiblities~\eqref{eq:susceptibilities_full_structure} in a real metasurface structure, with fully defined metaparticle material and shape, and with specific nonreciprocal elements. For the latter, we shall use here \emph{transistors}, for their advantages of spectral purity (single-frequency operation), symmetry-breaking low-cost source (DC battery) and biasing simplicity (DC circuit). Moreover, we shall consider a normal-incidence ($\theta = 0$) design, for simplicity, but the proposed procedure and structure are easily extensible to the case of oblique incidence.

\begin{figure}[h!]
  \centering
  \includegraphics[width=0.6\columnwidth]{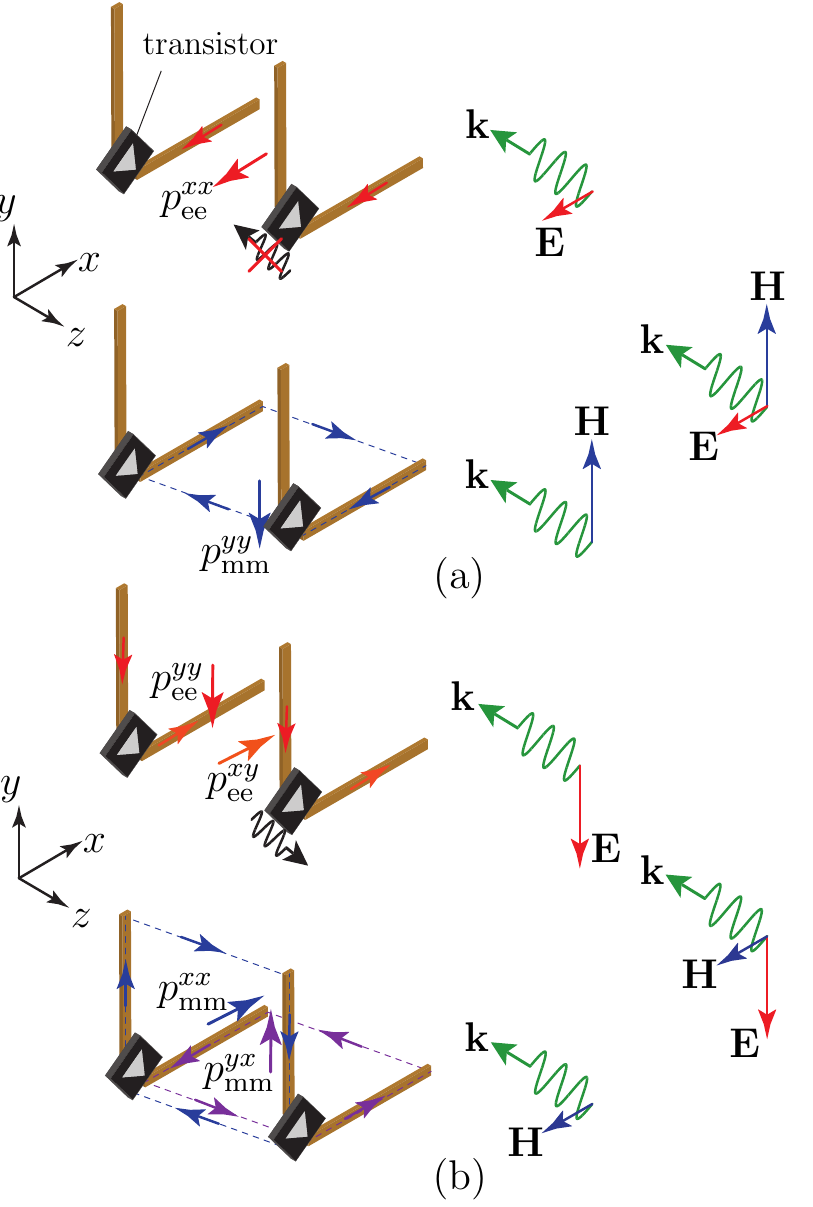}
  \caption{Proposed 2-layer transistor-loaded metaparticle to realize the susceptibilities in Eqs.~\eqref{eq:susceptibilities_full_structure} for the operation in~\figref{fig:generic_structure}. The notation $p_\text{ee}^{xy}$ represents the $x$ component of the electric dipole response due to the $y$ component of the electric field excitation, and so on.  (a)~$x$-polarization excitation. (b)~$y$-polarization excitation.}\label{fig:particle_2layer}
\end{figure}

We propose the 2-layer metaparticle implementation shown in~\figref{fig:particle_2layer} to realize the responses~\eqref{eq:susceptibilities_full_structure}. The metaparticle structure is composed of two identical L-shaped metal resonators, each loaded by a unilateral\footnote{In the case of a Field-Effect Transistor (FET), such a unilateral operation implies a common-source configuration, as typically used in RF amplifiers~\cite{Pozar_ME_2011}, whereas the common-gate configuration, typically used in logic electronics, is bilateral.} transistor chip at the corner of the L. The transistors are biased in the non-amplifying regime where they exhibit the ideal-isolator scattering response $\mathbf{S}_\text{tran}=[0,0;1,0]$, and they are oriented so that they pass currents flowing from the vertical section to the horizontal section and block currents flowing in the opposite direction.


Figure~\ref{fig:particle_2layer}(a) depicts the response of the metaparticle to an $x$-polarized wave. The $x$-direct incident electric field induces an electric dipole moment along the $x$ direction ($\chi_\text{ee}^{xx}$), without inducing any response along the $y$ direction due to transistor blocking ($\chi_\text{ee}^{yx} = 0$), while the $y$-directed incident magnetic field induces a magnetic dipole moment along the $y$ direction ($\chi_\text{mm}^{yy}$) without response along the $x$ direction ($\chi_\text{mm}^{xy}=0$). Figure~\ref{fig:particle_2layer}(b) depicts the response of the metaparticle to a $y$-polarized wave. In this case, the $y$-directed incident electric field induces electric dipole moments along both the $y$ \emph{and} $x$ directions ($\chi_\text{ee}^{yy}$ and $\chi_\text{ee}^{xy}$) via the current  passing across the transistor and, similarly, the $x$-directed incident magnetic field induces magnetic dipole moments along both the $x$ \emph{and} $y$ directions ($\chi_\text{mm}^{xx}$ and $\chi_\text{mm}^{yx}$). Hence, this configuration precisely provides the required non-zero and zero susceptibility components in~\eqref{eq:susceptibilities_full_structure}.

\begin{figure}[h!]
  \centering
  \includegraphics[width=0.75\columnwidth]{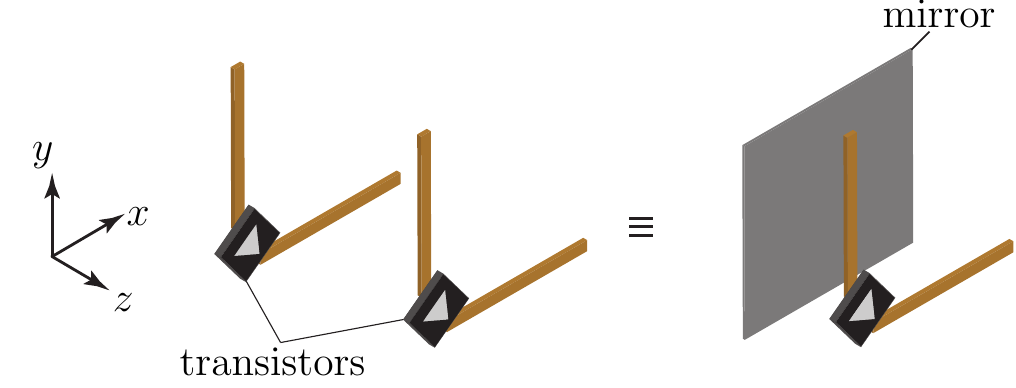}
  \caption{Equivalence in the $z>0$ (reflection) half-plane, according to image theory, between the initial metaparticle in \figref{fig:particle_2layer}, with structure recalled at the left, and the simpler mirror-backed structure, shown at the right.}\label{fig:image_equivalency}
\end{figure}

By symmetry, the metastructure in~\figref{fig:particle_2layer} is in fact equivalent, on the reflection side of the metasurface, to the simpler structure where the back resonator is suppressed and replaced by a mirror placed halfway between the two initial layers, as shown in~\figref{fig:image_equivalency}. Indeed, the latter structure, according to the image equivalence principle, exhibits the same scattering response as the former one. Given its greater simplicity, involving only one structured layer and only half the number of transistors, we shall adopt here this configuration.

\subsection{Metaparticle Design}\label{sec:metaparticle_design}

In the selected back-mirror metaparticle (right side in Fig.~\ref{fig:image_equivalency}), the design task reduces to determining the layer at top of the mirror. This layer represents a metasurface per se, which is different from the overall effective metasurface that it forms with the mirror, and this layer will therefore be subsequently considered as an independent metasurface, on top of a mirror-backed substrate.

\begin{figure}[h!]
  \centering
  \includegraphics[width=0.5\columnwidth]{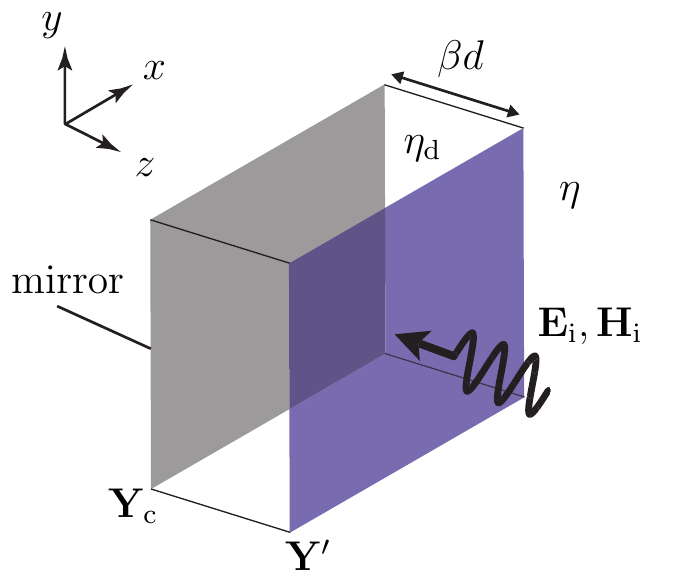}
  \caption{Admittance model for the mirror-backed structure in~\figref{fig:image_equivalency}.}\label{fig:admittance_model}
\end{figure}

In order to account for the multiple scattering occurring between the top metasurface and the mirror, we shall use the transmission-line model~\cite{pfeiffer2014bianisotropic} shown in~\figref{fig:admittance_model}. The metasurface and the mirror are modelled by the admittance matrices $\mathbf{Y}'$ and $\mathbf{Y}_\text{c}$, respectively, and are separated by a substrate of wave impedance $\eta_\text{d}$ and thickness $d$. The admittance matrix of the metasurface, whose parameters are to be determined, may be written as
\begin{equation}\label{eq:Y1_admitance}
  \mathbf{Y}' = \begin{bmatrix}
                     Y^{xx\prime} & Y^{yx\prime} \\
                     Y^{yx\prime} & Y^{yy\prime}
                    \end{bmatrix},
\end{equation}
while the admittance of the mirror, which will be realized by a simple conducting copper plate, is given by
\begin{equation}\label{eq:Y_2}
  \mathbf{Y}_\text{c} = i \sigma \mathbf{I},
\end{equation}
where $\sigma$ is the conductivity of the mirror, with $\sigma= 5 \times 10^7~1/\Omega$.

The transmission or ABCD matrix of the overall structure in \figref{fig:admittance_model} is then easily found by as
\begin{equation}\label{eq:admittance_model}
  \begin{bmatrix}
    \mathbf{A} & \mathbf{B} \\
    \mathbf{C} & \mathbf{D}
  \end{bmatrix} = \begin{bmatrix}
           \mathbf{I} & 0 \\
           \mathbf{n} \mathbf{Y}' & \mathbf{I}
         \end{bmatrix}
         \begin{bmatrix}
           \mathbf{I} \cos \beta d & -\mathbf{n} i \eta_d \sin \beta d  \\
           \mathbf{n} \frac{i \sin \beta d }{\eta_d} & \mathbf{I} \cos \beta d
         \end{bmatrix}
         \begin{bmatrix}
           \mathbf{I} & 0 \\
           \mathbf{n} \mathbf{Y}_\text{c} & \mathbf{I}
         \end{bmatrix},
\end{equation}
where
\begin{equation}\label{eq:I_n_deff}
  \mathbf{I} = \begin{bmatrix}
               1 & 0 \\
               0 & 1
             \end{bmatrix} \quad\text{and}\quad
\mathbf{n} = \begin{bmatrix}
              0 & -1 \\
              1 & 0
            \end{bmatrix}
\end{equation}
are the identity matrix and the 90$^\circ$-rotation matrix, respectively. The transmission matrix~\eqref{eq:admittance_model} can then be converted into its scattering matrix counterpart as~\cite{Pozar_ME_2011}

\begin{equation}\label{eq:abcd_to_s}
  \mathbf{S} = \begin{bmatrix}
        \mathbf{S}_{11} & \mathbf{S}_{12}  \\
        \mathbf{S}_{21}  & \mathbf{S}_{22}
      \end{bmatrix} =
      \begin{bmatrix}
        -\mathbf{I} & \frac{\mathbf{B} \mathbf{n}}{\eta}+\mathbf{A} \\
        \frac{\mathbf{n}}{\eta} & \frac{\mathbf{D} \mathbf{n}}{\eta}+\mathbf{C}
      \end{bmatrix}^{-1}
      \begin{bmatrix}
        \mathbf{I} & \frac{\mathbf{B} \mathbf{n}}{\eta}-\mathbf{A} \\
        \frac{\mathbf{n}}{\eta} & \frac{\mathbf{D} \mathbf{n}}{\eta}-\mathbf{C}
      \end{bmatrix},
\end{equation}
whose $\mathbf{Y}'$ (unknown) and other structural dependencies are naturally available from~\eqref{eq:admittance_model}.

For the mirror-backed metasurface structure to realize the operation in \figref{fig:generic_structure}, its reflection block, $\mathbf{S}_{11}$ in~\eqref{eq:abcd_to_s}, must equal the reflection matrix $\mathbf{R}_\text{spec}^\text{RGSI}$ in~\eqref{eq:desired_reflection_matrix}, and hence $(\mathbf{S}_\text{spec}^{\text{RGSI}})^\text{T}$ in~\eqref{eq:desired_reflection_matrix}. Enforcing this equality and solving for $\mathbf{Y}'$ yields
\begin{subequations}\label{eq:Y_1_value}
  \begin{equation}
      Y^{xx\prime} = Y^{yy\prime} = \frac{\eta^2 \sin(\beta d)-\eta_\text{d}^2 \sin (\beta d) + i \eta \eta_\text{d}\sigma \cos (\beta d)}{\eta_\text{d}^2 \sigma \sin (\beta d) - \eta_\text{d}^2 \sin (\beta d) + i \eta \eta_\text{d} \cos (\beta d)},
  \end{equation}
    \begin{equation}
      Y^{xy\prime} = 2  A e^{i \phi},
  \end{equation}
    \begin{equation}
      Y^{yx\prime} = 0.
  \end{equation}
\end{subequations}

To translate this admittance matrix into metasurface susceptibilities, we write the ABCD matrix corresponding to $\mathbf{Y}'$ as
\begin{equation}\label{eq:admittance_model_Y1_only}
  \begin{bmatrix}
    \mathbf{A}' & \mathbf{B}' \\
    \mathbf{C}' & \mathbf{D}'
  \end{bmatrix} = \begin{bmatrix}
           \mathbf{I} & 0 \\
           \mathbf{n} \mathbf{Y}' & \mathbf{I}
         \end{bmatrix}.
\end{equation}
convert this matrix to its scattering counterpart by reusing the formula~\eqref{eq:abcd_to_s}, and map this matrix, which we shall call $\mathbf{S}'=[\mathbf{S}_{11}',\mathbf{S}_{12}';\mathbf{S}_{21}',\mathbf{S}_{22}']$, to the surface susceptibility matrix according to the procedure that is described in~\cite{achouri2018design,achouri2020electromagnetic}, and that leads to the following equation:
\begin{subequations}
\begin{equation}\label{eq:S_to_X_matrix}
    \overline{\overline{\Delta}} = \Tilde{\overline{\overline{\chi}}}' \cdot \overline{\overline{A}},
\end{equation}
where
\begin{equation}\label{eq:mapb}
    \overline{\overline{\Delta}} = \begin{bmatrix}
      -\mathbf{m}/\eta+\mathbf{m} \mathbf{S}_{11}'/\eta+\mathbf{m} \mathbf{S}_{21}'/\eta & -\mathbf{m}/\eta+\mathbf{m} \mathbf{S}_{12}'/\eta+\mathbf{m} \mathbf{S}_{22}'/\eta \\
      -\mathbf{n} \mathbf{m} - \mathbf{n} \mathbf{m} \mathbf{S}_{11}' + \mathbf{n} \mathbf{m} \mathbf{S}_{21}' & \mathbf{n} \mathbf{m} - \mathbf{n} \mathbf{m} \mathbf{S}_{12}' + \mathbf{n} \mathbf{m} \mathbf{S}_{22}'
    \end{bmatrix},
\end{equation}
\begin{equation}\label{eq:mapc}
    \overline{\overline{A}} = \frac{1}{2} \begin{bmatrix}
      \mathbf{I} + \mathbf{S}_{11}' + \mathbf{S}_{21}' &\mathbf{I} + \mathbf{S}_{12}' + \mathbf{S}_{22}' \\
      \mathbf{n}/\eta - \mathbf{n} \mathbf{S}_{11} /\eta + \mathbf{n} \mathbf{S}_{21}'/\eta & -\mathbf{n}/\eta - \mathbf{n} \mathbf{S}_{12}' /\eta + \mathbf{n} \mathbf{S}_{22}'/\eta
    \end{bmatrix}
\end{equation}
and
\begin{equation}\label{eq:susceptibilities_normalized}
   \Tilde{\overline{\overline{\chi}}}'
   = \begin{bmatrix}
  -i \omega \epsilon \chi_\text{ee}^{xx\prime} & -i \omega \epsilon \chi_\text{ee}^{xy\prime} & -i k \chi_\text{em}^{xx\prime} & -i k \chi_\text{eme}^{xy\prime} \\
  i \omega \epsilon \chi_\text{ee}^{yx\prime} & i \omega \epsilon \chi_\text{ee}^{yy\prime} & i k \chi_\text{em}^{yx\prime} & i k \chi_\text{em}^{yy\prime} \\
 i k  \chi_\text{me}^{xx\prime} & i k \chi_\text{me}^{xy\prime} & i \omega \mu \chi_\text{mm}^{xx\prime} & i \omega \mu \chi_\text{mm}^{xy\prime} \\
  -i k \chi_\text{me}^{yx\prime} & -i k \chi_\text{me}^{yy\prime} & - i \omega \mu \chi_\text{mm}^{yx\prime} & -i \omega \mu \chi_\text{mm}^{yy\prime}
\end{bmatrix},
\end{equation}
with
\begin{equation}\label{eq:N1}
  \mathbf{m} = \begin{bmatrix}
                   1 & 0 \\
                   0 & -1
                 \end{bmatrix}.
\end{equation}
\end{subequations}

Substituting $\mathbf{S}'$ into~\eqref{eq:mapb} and~\eqref{eq:mapc}, inserting the resulting expressions into~\eqref{eq:S_to_X_matrix}, and inverting the resulting system yields the explicit susceptibility solutions, corresponding to~\eqref{eq:susceptibilities_normalized}:
\begin{subequations}\label{eq:susceptibilities_top_layer}
  \begin{equation}\label{eq:susceptibilities_top_layera}
    \chi_\text{ee}^{xx\prime}  =  \chi_\text{ee}^{yy\prime}  = \frac{i}{\omega \epsilon}\frac{\eta  \alpha i B  -\eta_\text{d} \gamma (\sigma-1) B}{\eta_\text{d}^3 \gamma^2 +\eta_\text{d}^3 \sigma^2 \gamma^2 - 2 \eta_\text{d}^3 \sigma \gamma^2},
\end{equation}
\begin{equation}\label{eq:susceptibilities_top_layerb}
\chi_\text{ee}^{xy\prime} = \frac{i}{\omega \epsilon} 2 A e^{i \phi},
\end{equation}
\begin{equation}\label{eq:susceptibilities_top_layerc}
\chi_\text{ee}^{yx\prime} = 0,
\end{equation}
\end{subequations}
where $\alpha = \cos (\beta d)$, $\gamma = \sin (\beta d)$, $B = (\eta^2 \gamma - \eta_\text{d}^2 \gamma + \eta_\text{d}^2 \sigma \gamma + \eta \eta_\text{d} \sigma \alpha i)$. The other susceptibility tensors, $\overline{\overline{\chi}}_\text{mm}$, $\overline{\overline{\chi}}_\text{em}$ and $\overline{\overline{\chi}}_\text{me}$, are found to zero, which reveals that the effective tensor $\overline{\overline{\chi}}_\text{mm}$, required from~\eqref{eq:susceptibilities_full_structureb}, is automatically provided by $xz$-loops formed between the top metasurface and the mirror, hence simplifying the former to a purely electrical homoanisotropic metasurface, characterized by the sole $\overline{\overline{\chi}}_\text{ee}$ susceptibility tensor.

\begin{figure}[h!]
  \centering
  \includegraphics[width=0.6\columnwidth]{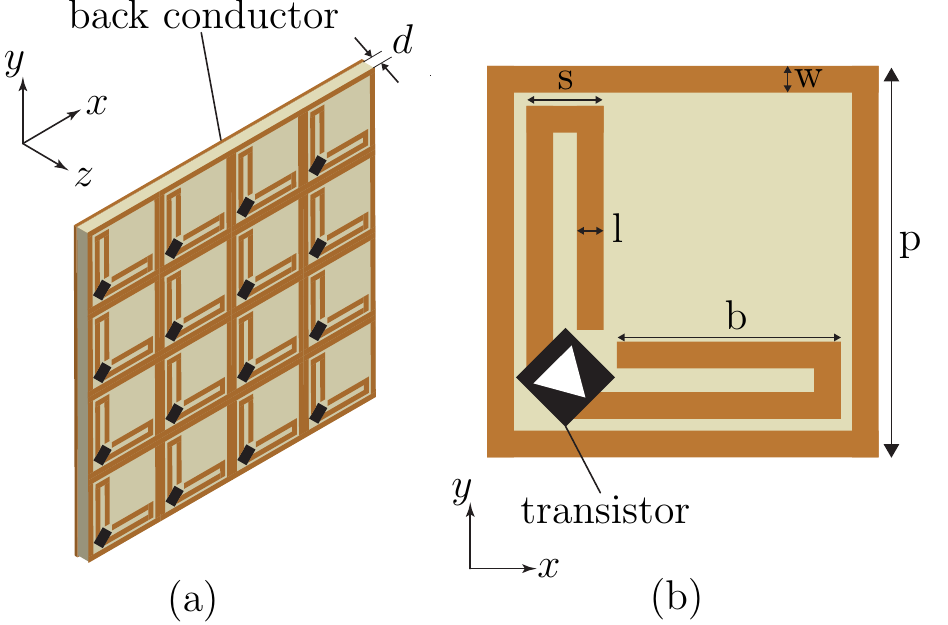}
  \caption{Proposed practical implementation of the RGSI metasurface. (a)~Perspective view. (b)~Front view of the unit cell.}\label{fig:scattering_particle}
\end{figure}

The last step of the design is to perform geometrical-parameter full-wave simulation mapping, as described in~\cite{achouri2020electromagnetic}. Figure~\ref{fig:scattering_particle} shows the final metaparticle design, where we folded the strips into C-section structure for better subwavelength confinement. Note that the currents in the parallel strips of the C-sections do not fully cancel out due to resonance non-uniformity (zero current at the edges and maximum at the center of the unfolded strip structure), which provides the same responses as those previously described despite the smaller footprint.

\section{Results}\label{sec:results}

This section present full-wave simulation results, using the commercial software CST Microwave Studio, for the RGSI metasurface implementation in \figref{fig:scattering_particle}. We shall consider and compare two implementations: one using quasi-ideal unity-gain unilateral transistors, corresponding to a quasi-ideal isolator with scattering matrix $\mathbf{S}_\text{trans}=[0,I;1,0]$, with $I=-30$~dB, and one using the HMC441LP3E transistor chip from Analog Devices, with (frequency dependent)  scattering parameters, including gain, given in the data sheet of the chip provided by the company. The design frequency is set to 7.5~GHz. For both implementations, we shall plot the simulated scattering parameters versus frequency, and the susceptibilities of the top layer extracted from these scattering parameters for comparison with the ideal susceptibilities given by~\eqref{eq:susceptibilities_top_layer}\footnote{This extraction is done, following the method described in~\secref{sec:metaparticle_design}, as follows: 1)~equating the simulated scattering matrix to the scattering matrix~\eqref{eq:abcd_to_s} with~\eqref{eq:admittance_model}, 2)~solving the resulting equations for the admittance matrix $\mathbf{Y}'$, and 3)~translating this so-obtained admittance matrix into surface susceptibilities using~\eqref{eq:admittance_model_Y1_only} and~\eqref{eq:S_to_X_matrix}.}.

\begin{figure}[h!]
  \centering
  \includegraphics[width=0.65\columnwidth]{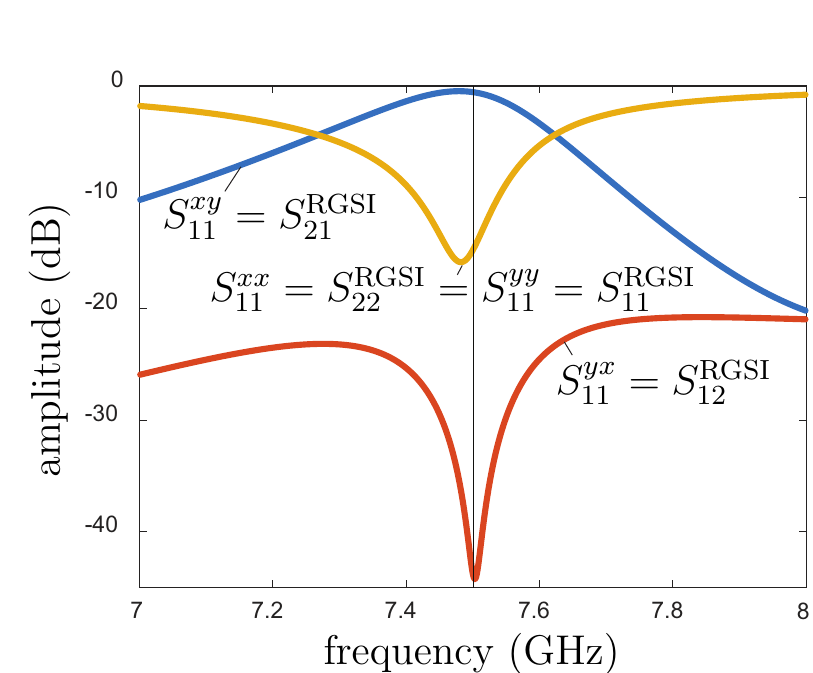}
  \caption{Full-wave simulated scattering parameters for the RGSI metasurface in~\figref{fig:scattering_particle} using gain-less unilateral transistor, with a substrate of $\epsilon_\text{r} = 6.2$ and for the parameters $l=0.5$~mm, $d=3$~mm, $w=1$~mm, $b=9$~mm, $s=1.5$~mm, and $p=14.8$~mm.}\label{fig:sparam_ideal_isolator}
\end{figure}

\begin{figure}[h!]
  \centering
  \includegraphics[width=0.75\columnwidth]{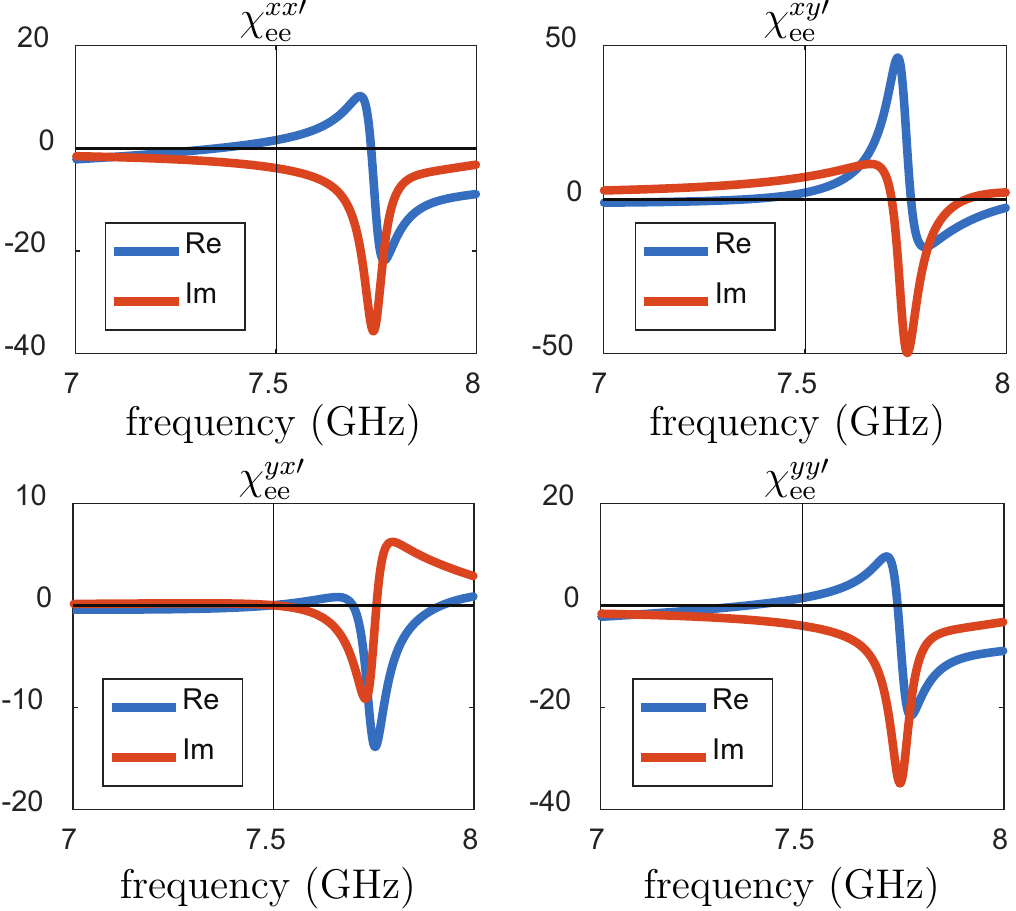}
  \caption{Electric susceptibilities for the top metasurface layer in~\figref{fig:scattering_particle}, extracted from the scattering parameters in~\figref{fig:sparam_ideal_isolator}.}\label{fig:extracted_X_ideal}
\end{figure}

Figures~\ref{fig:sparam_ideal_isolator} and~\ref{fig:extracted_X_ideal} present the results for the RGSI metasurface with the quasi-ideal unity-gain unilateral transistor. The desired RGSI operation (\figref{fig:generic_structure}) is clearly observed at the design frequency (7.5~GHz) in \figref{fig:sparam_ideal_isolator}, where the metasurface exhibits and isolation of around $40$~dB between the cross-polarized ports $S_{11}^{xy}$ and $S_{11}^{yx}$, and a matching of $-15$~dB for both co-polarized reflections. Moreover, the design results $\chi_\text{ee}^{xx\prime} = \chi_\text{ee}^{yy\prime}$ of~\eqref{eq:susceptibilities_top_layera}, $\chi_\text{ee}^{xy\prime}\neq 0$ of~\eqref{eq:susceptibilities_top_layerb} and $\chi_\text{ee}^{yx\prime}=0$ of~\eqref{eq:susceptibilities_top_layerc}, satisfying the nonreciprocity relation $\chi_\text{ee}^{yx\prime}\neq\chi_\text{ee}^{xy\prime}$, are verified in \figref{fig:extracted_X_ideal}.

\begin{figure}[h!]
  \centering
  \includegraphics[width=0.6\columnwidth]{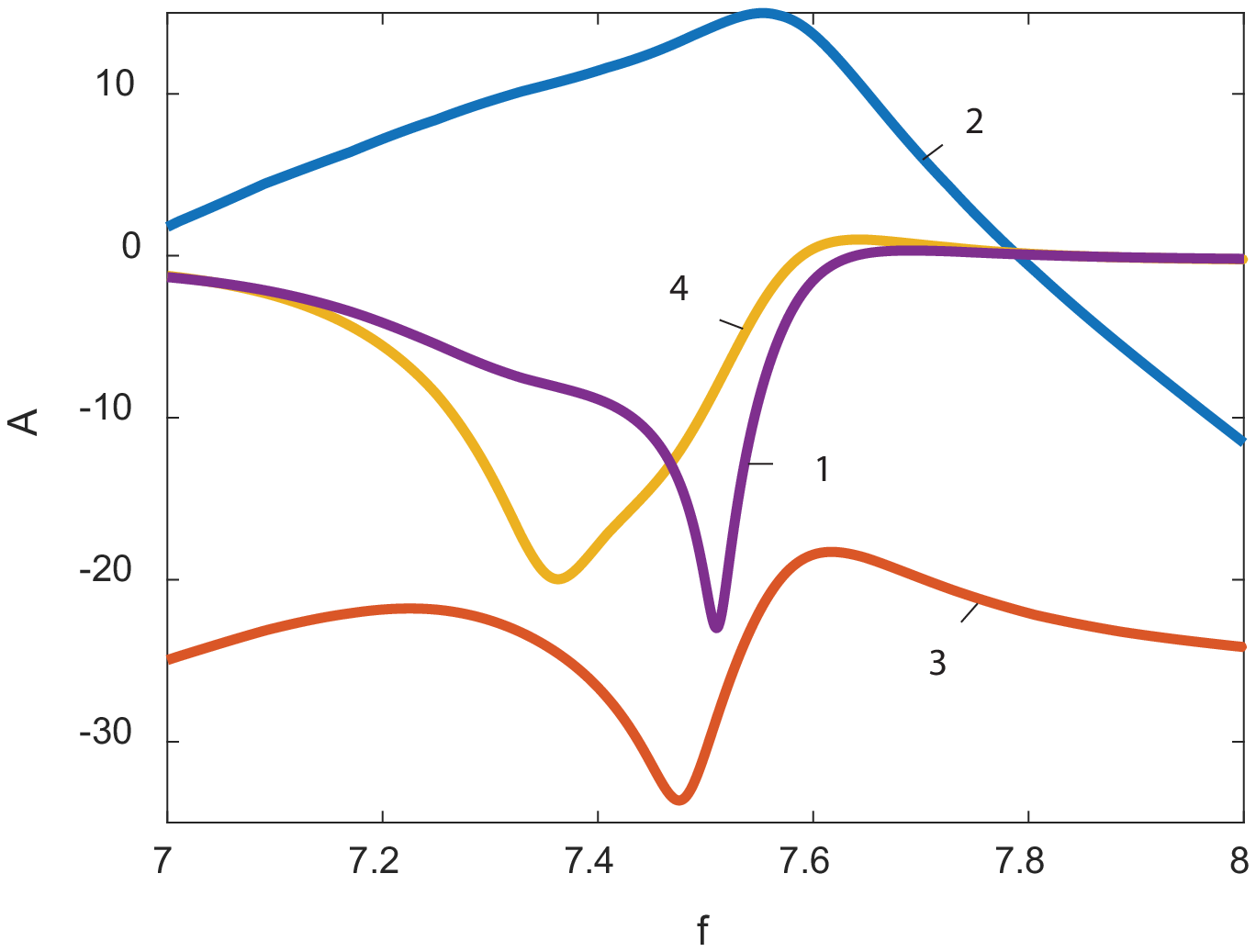}
  \caption{Full-wave simulated scattering parameters of the RGSI metasurface of~\figref{fig:scattering_particle} using the HMC441LP3E transistor chips, and for the same substrate and geometric parameters as in \figref{fig:sparam_ideal_isolator}.}\label{fig:sparam_amplifier}
\end{figure}

\begin{figure}[h!]
  \centering
  \includegraphics[width=0.75\columnwidth]{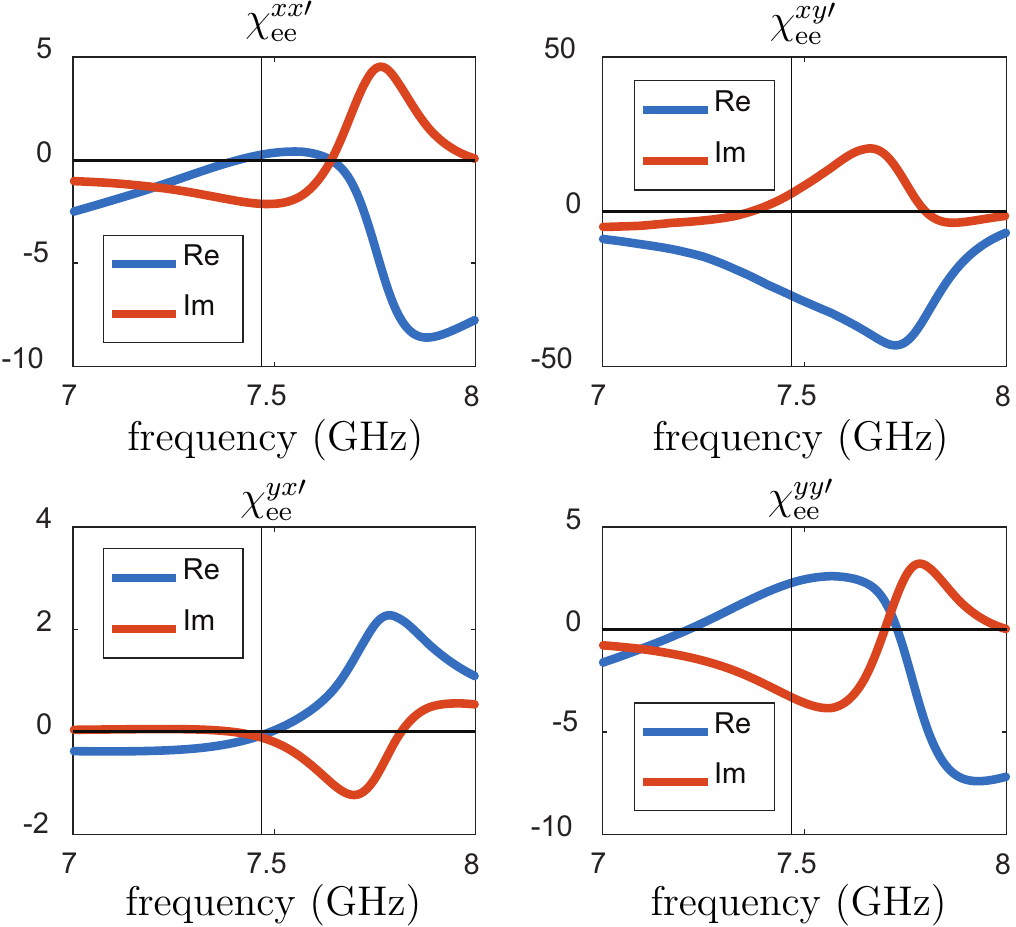}
  \caption{Electric susceptibilities for the top layer, extracted from the susceptibilities in~\figref{fig:sparam_amplifier}.}\label{fig:extracted_X_amplifier}
\end{figure}

Figures~\ref{fig:sparam_amplifier} and~\ref{fig:extracted_X_amplifier} present the results for the RGSI metasurface with the HMC441LP3E transistor chips. The spectrum observed in \figref{fig:sparam_amplifier} is slightly different from the design target, due to asymmetries of the chip that were not accounted for in the design; a better operation frequency here could be $7.467$~GHz, which features the best trade-off between gain, isolation and matching. At this frequency, a good RGSI operation is achieved, with a gain of $13$~dB ($S_{11}^{xy}$), an isolation of over $40$~dB (with respect to $S_{11}^{yx}$) and equal port matching of $-12.9$~dB ($S_{11}^{xx}$ and $S_{11}^{yy}$). The extracted susceptibilities in \figref{fig:sparam_amplifier}, although quite different from those obtained for the uasi-ideal unilateral transistors (in~\figref{fig:extracted_X_ideal}), still satisfy $\chi_\text{ee}^{xy\prime}\neq 0$ of~\eqref{eq:susceptibilities_top_layerb} and $\chi_\text{ee}^{yx\prime}=0$ of~\eqref{eq:susceptibilities_top_layerc}, whereas the relation~\eqref{eq:susceptibilities_top_layera} is not satisfied anymore, due the asymmetry of the transistor chip, fortunately without fatal consequence on the RGSI operation of the metasurface, as we saw in \figref{fig:sparam_amplifier}

\section{Conclusion}\label{sec:conclusion}
We have presented the concept of a magnetless RGSI metasurface. We have derived the surface susceptibility tensors required to realize this operation, and proposed a transistor-based mirror-backed implementation of a corresponding RGSI metasurface. Finally, we have demonstrated the device by full-wave simulations for both quasi-ideal unity-gain isolators and commercial  transistor chips with gain. This RGSI metasurface may be used in various electromagnetic applications and as a step towards more sophisticated magnetless nonreciprocal systems.

\appendix
\section{Bianisotropic Design}\label{sec:bian_des}

In~\secref{sec:suscepbility_synthesis}, we made the homoanisotropic choice of the 8 susceptiblity components $\chi_\text{ee}^{xx}$, $\chi_\text{ee}^{xy}$, $\chi_\text{ee}^{yx}$, $\chi_\text{ee}^{yy}$, $\chi_\text{mm}^{xx}$, $\chi_\text{mm}^{xy}$, $\chi_\text{mm}^{yx}$ and $\chi_\text{mm}^{y}$ to implement the proposed magnetless gyrotropic reflective spatial isolator metasurface, but we could have chosen a different set of eight susceptibility components.

Let us consider here the alternative axial bianisotropic set $\chi_\text{ee}^{xx}$, $\chi_\text{ee}^{yy}$, $\chi_\text{mm}^{xx}$, $\chi_\text{mm}^{yy}$, $\chi_\text{em}^{xx}$, $\chi_\text{em}^{yy}$, $\chi_\text{me}^{xx}$, $\chi_\text{me}^{yy}$, where the  gyrotropic components are now $\chi_\text{em}^{xx}$, $\chi_\text{em}^{yy}$, $\chi_\text{me}^{xx}$ and $\chi_\text{me}^{yy}$ instead of $\chi_\text{ee}^{xy}$, $\chi_\text{ee}^{yx}$, $\chi_\text{mm}^{xy}$ and $\chi_\text{mm}^{yx}$. Following the same procedure as in~\secref{sec:suscepbility_synthesis} for this alternative set yields
\begin{subequations}\label{eq:susceptibilities_bianisotropic}
  \begin{equation}
    \overline{\overline{\chi}}_\text{ee} = \begin{bmatrix}
                                             \chi_\text{ee}^{xx} & \chi_\text{ee}^{xy} \\
                                             \chi_\text{ee}^{yx} & \chi_\text{ee}^{yy}
                                           \end{bmatrix} = \begin{bmatrix}
                                                             \frac{-2i \sec \theta}{k} & 0 \\
                                                             0 &  \frac{-2i \cos \theta}{k}
                                                           \end{bmatrix},
  \end{equation}
    \begin{equation}
    \overline{\overline{\chi}}_\text{mm} = \begin{bmatrix}
                                             \chi_\text{mm}^{xx} & \chi_\text{mm}^{xy} \\
                                             \chi_\text{mm}^{yx} & \chi_\text{mm}^{yy}
                                           \end{bmatrix} = \begin{bmatrix}
                                                             \frac{-2i \sec \theta }{k} & 0 \\
                                                             0 &  \frac{-2i \cos \theta }{k}
                                                           \end{bmatrix},
  \end{equation}
      \begin{equation}
    \overline{\overline{\chi}}_\text{em} = \begin{bmatrix}
                                             \chi_\text{em}^{xx} & \chi_\text{em}^{xy} \\
                                             \chi_\text{em}^{yx} & \chi_\text{em}^{yy}
                                           \end{bmatrix} = \begin{bmatrix}
                                                             \frac{-4i A e^{i \phi} \sec \theta }{k} & 0 \\
                                                             0 &  0
                                                           \end{bmatrix},
  \end{equation}
      \begin{equation}
    \overline{\overline{\chi}}_\text{me} = \begin{bmatrix}
                                             \chi_\text{me}^{xx} & \chi_\text{me}^{xy} \\
                                             \chi_\text{me}^{yx} & \chi_\text{me}^{yy}
                                           \end{bmatrix} = \begin{bmatrix}
                                                             0 & 0 \\
                                                             0 &  \frac{-4i A e^{i \phi} \cos \theta }{k}
                                                           \end{bmatrix}.
  \end{equation}
\end{subequations}

This alternative solution would naturally lead to different metaparticles than those used in the paper. Particularly, the magnetodielectric coupling terms would imply chiral, $z$-asymmetric metaparticles~\cite{Caloz_EC_PI_2019,Caloz_EC_PII_2019}.

\bibliographystyle{IEEEtran}
\bibliography{LIB}

\end{document}